# Radiation Exposure Theory
# Comparison of data on Mutation Frequencies of Mice

Revised 4/20/2013  00:20:00

Yuichiro Manabe[1] and Masako Bando[2]

[1] Division of Sustainable Energy and Environmental Engineering,
Graduate School of Engineering, Osaka University,
2-1 Yamada-oka, Suita-shi, Osaka, Japan, 565-0871
e-mail address: manabe_y@see.eng.osaka-u.ac.jp

[2] Yukawa Institute for Theoretical Physics, Kyoto University,
Kitashirakawa oiwake-cho, Sakyo-ku, Kyoto-shi, Kyoto, Japan, 606-8502

We propose Radiation Exposure Theory (RET), a mathematical framework to estimate biological damage caused by irradiation. This is an extension of LDM model which was proposed in the paper [Y. Manabe et al.: J. Phys. Soc. Jpn. 81, 104004(2012)]. The theory is based on physical protocol, "a stimulus and its response''. It takes account of considerable response including mutation, cell death caused by outer stimulus, as well as biological functions such as proliferation, apoptosis and repair. By taking account of biological issues, namely a variety of preventable effects, which is characteristic feature of living object. RET can explain various data which simple LNT does not reproduce. As one of the characteristic features of RET, we propose a scaling law, namely all the data point with different dose rate irradiation are predicted to lie on the universal line if the variables, the biological damage function as a function of time development after artificial irradiation starts are converted to renormalized ones. The above scaling law can be compared with experimental data. We adopt the accumulated experiments performed by so-called mega mouse projects. It is found that their data points are converted into a universal scaling function and are in reasonably agreement with the prediction of RET.

**Keywords:** Mega-mouse project, radiation risk, apoptosis, DNA repair, mathematical model, low-dose exposure

## I. INTRODUCTION

It is one of the most controversial issues how irradiation, especially low dose one hurts biological objects. If it is merely a physical process, the frequency of radiation-induced mutations is naively thought to be proportional to its total dose. In most cases biological damage begins with the mutation of living cells which is caused by ionization. Such kind of ionization is usually induced in proportion to the energy deposit from radiation from outside. Indeed in 1927, Herman J. Muller studied the effect of X-rays on Drosophila [1], and showed a linear dependence on total artificial irradiation dose exposure to the number of mutation frequency, without any threshold effects. This observation caused a strong impact not only on the scientific community but on whole society and has led to official adoption of what is called "LNT hypothesis".

Later W. Russell, of the Oak Ridge National Laboratory, proposed to test its validity in mice by using huge number of mice (7 million mice) and got important data on mutation frequency. Those famous experimental data are summarized by Russell and Kelly [2]. This provides us with important information on the frequencies of transmitted specific-locus mutations induced by artificial irradiation in mouse spermatogonia stem-cells. Their results indicate that the level of dose rate gives important effects to the mutation frequency, which indicates the existence of mechanisms by which cells can be protected against irradiation. The result of this mega mouse project is still the most famous and frequently referred by many authors.

However the arguments is not enough to clarify the quantitative estimation: they just divides the data into 2 groups, high and low dose rates and determine the slopes of linear dependence of mutation frequency vs. total dose of each group. Many arguments to support LNT as well as those insisting lower (hormesis effects, for example) or higher (coming from bystander effects) risk than LNT. This is because the subsequent biological processes make sometimes preventable effects on the one hand, and on the other hand yield enhancement of mutation of living cells.

We have to make quantitative estimation of various kinds of mechanism operating on living objects, and it is indeed necessary to construct a framework to make systematic analysis taking account of every possible effect in order to make quantitative estimation.

In a separate paper we propose a model to estimate biological damage caused by radiation, which we call LDM

(Low Dose Meeting) Model [3]. However the original LDM model is not always applicable to the case of more general situation, although it shows how the recovery effects suppress the rapid proliferation of broken cells in realistic living objects.

The aim of this paper is to formulate a model by taking account of possible conceivable biological effects under more general situations in living objects. We take a most reasonable approach based on "a stimulus and its response" and formulate a set of differential equations for the numbers of normal and broken cells in a system of a biological object. Hereafter we call this simply RET (Radiation Exposure Theory). We here take the most characteristic feature predicted from RET and compare the most famous experiments known as "Mega-mouse project" with our prediction of "scaling law" [2, 4].

It is found that the numerical results of the mutation frequency of mice are in reasonable agreement with the experimental data, especially we can confirm that the scaling law works well.

In section II, we propose general formalism of RET by taking systematically account of all the effects including cell death, cell density dependence and so on. Section III is devoted to the study of the structure of RET, proposing a scaling law. As a typical example, in section IV we apply RET to the summary of experimental data obtained by Mega mouse project, and see how well they are on the predicted line by scaling law. Summary and future prospect are presented in section V.

## II. RET FORMULATION

Consider a system of cells, a tissue or an organ (hereafter we call it symbolically "tissue") with its cell number capacity $(N_n)_{max} = K$, the maximum number of normal cells of a tissue. FIG.1 shows a dynamical situation of stimulus-response of the cells in a system caused by irradiation (straight wave lines) together with other stimulus (dotted wave lines). In living object there exist various types of functions, proliferation, repair and apoptosis effects (denoted by straight lines). Further we take account of cell death effects caused by irradiation as well. The arrow for each line indicates input or output source.

Suppose that a tissue contains normal cells with its number $N_n$ and broken cells, $N_b$ and at $t = 0$ it is exposed by artificial radiation with dose-rate $d(t)$. Hereafter let the numbers, $N_n$ and $N_b$, be normalized by their maximum number of the system, $K$, which are denoted by $F_n(t) = \frac{N_n(t)}{K}$ and $F_b(t) = \frac{N_b(t)}{K}$, respectively. The risk estimate is to be related to those functions, the ratio of broken cells, which may turn to a cancer tumor, to normal ones.

The numbers of normal and broken cells which are usually proliferating, respond to the stimulus coming from environment, natural or artificial. Such situation is schematically expressed as in FIG. 1.

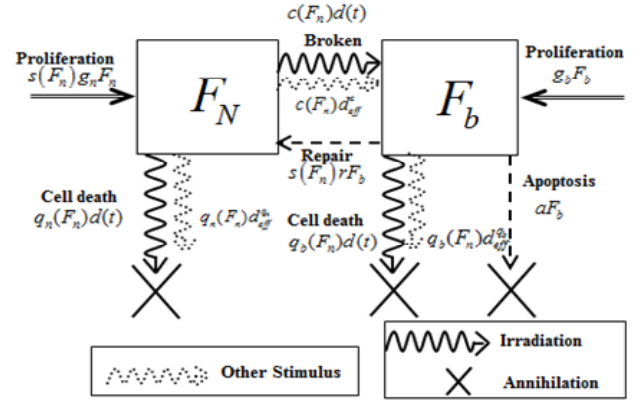

FIG. 1. Stimulus response diagram of RET.

The time dependence of number functions, $F_n(t), F_b(t)$, obeys according to their in and outcomes seen in FIG. 1;

$$\frac{d}{dt}F_n(t) = s(F_n)[g_n F_n(t) + r F_b(t)] - c(F_n)(d(t) + d_{eff}^c) - q_n(F_n) \cdot (d(t) + d_{eff}^{q_n}),$$

$$\frac{d}{dt}F_b(t) = c(F_n) \cdot (d(t) + d_{eff}^c) - q_b(F_b) \cdot (d(t) + d_{eff}^{q_b}) + g_b F_b(t) - (r + a) F_b(t). \quad (1)$$

The notations $g_n$, $g_b$, $r$ and $a$, are proliferation rate of normal, broken cells, the rates of inducing repair and apoptosis of broken cells, respectively.

Some comments are necessary in order to relate the parameters in Fig. 1 to Eq. (1).

The cell growth rates, $g_n$, $g_b$, correspond to the input contribution to $F_n(t)$, $F_b(t)$, which are proportional to, $F_n(t)$, $F_b(t)$, respectively. Note that we here introduce the suppression factor $s(F_n(t))$ to control the proliferation or growth of normal cells. This is because of the characteristic property of normal cells, namely they arrest their increase when the number of normal cells approaches to its maximum $K$, $N_n(t) \to K$,

$$\lim_{F_n \to 1} s(F_n(t)) \to 0,$$
$$s(F_n(t)) = (1 - F_n(t)). \qquad (2)$$

On the other hand, broken cells never stop their proliferation.

As another source of income to $F_n(t)$, the repair function takes broken cells into the group of normal cell group, which is corresponding to $rF_b(t)$ term in Eq. (1).

The output part of $F_n(t)$ consists of mainly two parts, both cause decrease of normal cells.

First, note that the term with breaking coefficients $c(c')$, representing transition of normal cell to broken one caused by outer stimulus; artificial irradiation or background. The coefficient $c(N_n)$ are called "radio sensitivity". It depends on what kind of tissue or organ, and what kind of living object we are investigating. It is a breaking coefficient of the irradiation strength rate $d(t)$. In general, the coefficient $c(c')$ are to be determined by radiation cross section of cells, cell density, and the related surrounding conditions, such as temperature, density of various kinds of proteins and so on. However once we fix the biological conditions of the system they are fixed parameters. Although the sensitivity parameter depends on the number of normal cells in general, let us remind that we here use the unit Gy for the strength of irradiation. This represents the unit of absorbed dose, the absorption of one joule of energy, inducing ionizing radiation, per kilogram of matter. Thus in terms of Gy, the irradiation strength rate $d(t)$ deposits a corresponding increment of energy per time in unit volume of tissue, and thus the derivative of total number of broken cells, is proportional to the amount of irradiation strength rate $d(t)$ only, so far as we consider the case where normal cells in a tissue is dense enough. Such kind of treatment may be similar to the situation in which nuclear physicists often employs the concept of nuclear matter which is defined as an idealized system consisting of a huge number of nucleons with finite density. Under such situation we can take $c(N_n)$ being independent of cell number function $F_n(t)$ unless it is extremely small. On the contrary if the cell density becomes very small, namely the number of normal cells is very small and the radioactive energy deposit is not fully poured into the breakdown of normal cells, $c(N_n)$ becomes proportional to $F_n(t)$. Therefore $c(F_n)$ should have its asymptotic form as follows,

$$\lim_{F_n \to 0} c(F_n(t)) = c'F_n(t), \quad \lim_{F_n \to 1} c(F_n(t)) = c. \qquad (3)$$

The other term, which is also proportional to the irradiation strength $d(t)$, leads normal cells into death, cell death term. In much the same manner we define the following coefficients as well,

$$\lim_{F_n \to 0} q_n(F_n(t)) = q'_n F_n(t), \lim_{F_n \to 1} q_n(F_n(t)) = q_n. \qquad (4)$$

As for the broken cells, the situation is almost the same except that we introduce the effect of apoptosis and repair terms which are proportional to the number function of broken cells $F_n(t)$ with coefficient $a$ and $r$ respectively. Since the parameters, $q_n$ and $q_b$, are the rates of death of normal and broken cells caused by artificial irradiation $d(t)$, respectively, the definition is much the same as for the normal cell side.

$$\lim_{F_b \to 0} q_b(F_b(t)) = q'_b F_b(t), \quad \lim_{F_b \to 1} q_b(F_b(t)) = q_b. \qquad (5)$$

A comment is in order on the stimulus from outside other than artificial irradiation. Living object are receiving stimulus from all amount of time-independent background surroundings, for which we denote $d^c_{eff}$ as a whole. The effective sum of dose equivalent strengths, $d^c_{eff}$ and $d^d_{eff}$ may be naively guessed to be equal, but in general they might not be the same because we do not know the origin of background stimulus, there might be physical, (including natural radiation) or chemical, biological or other unidentified sources. The equivalent background dose rate, $d_{eff}$, is in some sense the stimulus to which we convert the whole amount of those coming from those other than artificial irradiation. It can be taken as almost constant since it comes from the environment where a living object lives. Note that the stimulus from other sources breaks normal cells into broken ones from the very starting point of living objects (at birth of living object). Living objects are receiving such sort of stimulus from their surroundings just after they are born. Their effects may be balanced with their preventable effects such as repair and apoptosis. The terms coming from repair and apoptosis effects, which we can totally call "preventable effect" [5], however, usually overwhelms proliferation effect, unless living objects would have died.

For long time during they live, the number of normal and broken cells tends to certain numbers maintaining stationary state until artificial irradiation starts. Let us consider such realistic situation and take account of all the effects which we have to consider.

## III. STRUCTURE OF RET AND LNT

In order to see the characteristic feature of RET, let us study the behavior of $F_n(t)$, $F_b(t)$, by solving the differential equation of Eq.(1) for typical case as shown in FIG.2.

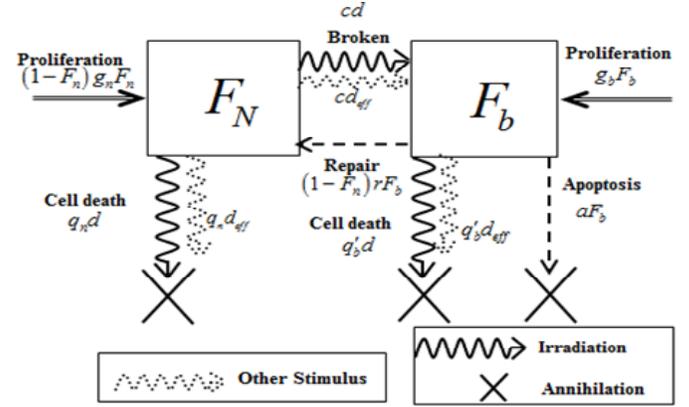

FIG. 2. Stimulus response diagram (typical example making simple approximation)

We shall see that the differential equation of $F_b(t)$ is analytically obtained for the case of time independent irradiation rate exposure.

Let us consider the case where an artificial irradiation with time independent irradiation rate, $d$, starts at $t=0$;

$$d(t) = \begin{cases} 0 & (t<0) \\ d & (t\geq 0) \end{cases}. \quad (6)$$

Then Eq. (6) can be simply expressed as,

$$\frac{d}{dt}F_n(t) = (1-F_n(t))[g_n F_n(t) + rF_b(t)] - c(F_n(t))(d+d_{eff}^c) - q(F_n(t))(d+d_{eff}^{q_n}),$$

$$\frac{d}{dt}F_b(t) = c(F_n)\cdot(d+d_{eff}^c) - q_b(F_b)\cdot(d+d_{eff}^{q_b}) + g_b F_b(t) - (r+a)F_b(t). \quad (7)$$

Note that the stimulus consists of two parts, the first term $d$ represents artificial irradiation which we are under consideration and the second term, total background stimulus representing various cues from the background environment other than artificial irradiation. Since the background stimulus comes from surrounding environment ever since the birth of a living object, the stimulus may be time independent on the average.

Let us see the solution of the above equation for the case where

$$c(F_n(t)) = c,\ q(F_n(t)) = q_n,\ q(F_b(t)) = q'_b F_b(t), \quad (8)$$

which corresponds to the situation where a system consists of almost normal cells with a small amount of broken cells, and normal cells can be treated in analogous to nucleons in nuclear matter with constant density. Then Eq. (7) is simply reduced to the following form,

$$\frac{d}{dt}F_n(t) = (1-F_n(t))[g_n F_n(t) + rF_b(t)] - c(d+d_{eff}^c) - q_n(d+d_{eff}^{q_n})$$
$$= (1-F_n(t))[g_n F_n(t) + rF_b(t)] - (c+q_n)d - (cd_{eff}^c + q_n d_{eff}^{q_n}),$$
$$\frac{d}{dt}F_b(t) = c(d(t)+d_{eff}^c) - \{q_b'(d(t)+d_{eff}^{q_b}) + r + a - g_b\}F_b(t). \quad (9)$$

The notations $d_{eff}$ represent various kinds of time independent stimulus from surroundings shown in Fig. 1 and their values are balanced with the parameters, $c, q_b', r, a, g_b$ of a system to keep the stationary condition,

$$\frac{d}{dt}F_b(t) = 0 \text{ with } d=0;$$

$$\frac{d}{dt}F_b(t) = cd_{eff}^c - \{q_b' d_{eff}^{q_b} + \mu\}\overline{F}_b = 0,$$

$$\Rightarrow \overline{F}_b = \frac{cd_{eff}^c}{q_b' d_{eff}^{q_b} + \mu}. \quad (10)$$

Here a comment is in order. The stationary state can be realized only when under the following condition,

$$q_b' d_{eff}^{q_b} + \mu > 0, \quad \mu = r + a - g_b. \quad (11)$$

It must be remarked that $\overline{F}_b$ in Eq. (10) is defined as so-called "control", implying that a living object always keeps a certain number of broken cells inside so far as it receives a certain stimulus from outside. This is naturally understood since it lives long in real world. Also note the condition Eq. (11) implies that the total contribution of repair, apoptosis and cell death effects should exceed the proliferation term in order for the system to become stationary.

Let the solve the equation for $F_b(t)$ of Eq. (9), which we express simply as,

$$\frac{d}{dt}F_b(t) = A - BF_b(t),$$
$$A \equiv c(d+d_{eff}^c), B \equiv q_b'(d+d_{eff}^{q_b}) + \mu, \quad \mu \equiv r+a-g_b. \quad (12)$$

Relating to the above discussion, we first classify the situations into three cases,

$$\begin{cases} \text{case1} & B=0, \\ \text{case2} & B<0, \\ \text{case3} & B>0, \\ B \equiv q_b'(d+d_{eff}^{q_b}) + \mu, \quad \mu \equiv r+a-g_b. \end{cases} \quad (12)$$

For the case $B=0$, which may be specific case, in which the total dose is just arranged to cancel the total amount of preventable effects, the solution is

$$F_b(t) = c(d+d_{eff}^c)t + F(0) = c(D+D_{eff}^c) + F(0), D \equiv dt, D_{eff}^c \equiv d_{eff}^c t. \quad (13)$$

The notations $D(D_{eff})$ are irradiated total dose up to time t. This shows just LNT behavior. If we arrange the experimental condition in such a way that no preventable effect works, we can get the specific experimental data of such LNT behavior. One example of such situation is the case of X-rays induced lethal mutations on Drosophila performed by Herman J. Muller [1]. No preventable effects is known to work for the system of spermatogonial cells, showing a linear dependence on total artificial irradiation dose exposure to the number of lethal mutation frequency, without any threshold effects.

Another situation might be accidentally realized if we breed P53 genetic knockout mice under a certain ideal circumstance without any stimulus, for example.

The case $B<0$ may correspond to the case where the proliferation power is too strong and overwhelms the amount of preventable effects, so the number function of broken cells exceeds LNT and is growing up rapidly. There might exist specific tissues or organ in nature to have such characteristics.

In usual cases, living objects inevitably acquire preventable function during a long history of evolution, we can take $B>0$, yielding to the following solution,

$$\frac{d}{dt}F_b(t) = A - BF_b(t), B > 0$$

$$\Rightarrow F_b(t) = \left(\frac{A}{B} - F_b(0)\right)(1 - \exp(-Bt)) + F_b(0),$$

$$A \equiv c(d + d_{eff}^c), B \equiv q_b'(d + d_{eff}^{q_b}) + \mu, \mu \equiv r + a - g_b. \quad (14)$$

The initial value is equal to $\bar{F}_b$ because this corresponds to the control in experiments just before artificial irradiation starts, i.e.,

$$F_b(0) = \bar{F}_b = \frac{cd_{eff}^c}{q_b'd_{eff}^{q_b} + \mu}. \quad (15)$$

Now that we saw global structure of the solutions for $F_b(t)$, let us show in Fig. 3 typical examples of $F_b(t)$ to visualize the difference of behavior for three cases.

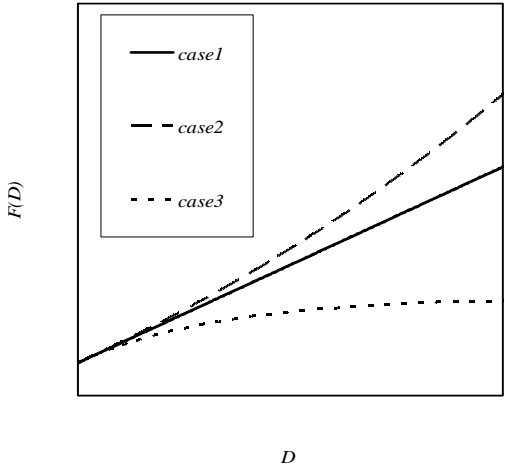

FIG. 3. Examples for visualizing the difference of behavior for three cases showing LNT (B=0) and more (B<0) and less (B>0).

Here we adopt total the total dose exposure as a horizontal axis by converting time dependence into total dose dependence since total dose in this case is proportional to total dose as is defined in Eq. (13).

If we further make simple assumption that the equivalent dose rate acting to mutant and death terms are equal, $d_{eff}^{q_n} = d_{eff}^c \equiv d_{eff}$, Eq. (15) can be reduced to simple form,

$$\bar{F}_b = \frac{cd_{eff}}{q_b'd_{eff} + \mu}, \quad (16)$$

implying that the background stimulus is obtained uniquely from the experimental value of control; $d_{eff}$ is also determined from the control initial value $F_b(0)$. This can be read off from the experimental data. Thus the above function is expressed in terms of the given three parameters, $c, q_b', \mu$ which are to be fixed so as to reproduce experimental data. Of course we fix the set up experimental situation of given dose rate $d$ as well.

Now that we have time development of the number function of broken cells, $F_b(t)$, the one of normal cells $F_n(t)$ which obeys the following equation,

$$\frac{d}{dt}F_n(t) = [1 - F_n(t)][g_nF_n(t) - q_nF_n(t) + rF_b(t)] - c(d + d_{eff})$$
$$= [1 - F_n(t)][(g_n - q_n)F_n(t) + rF_b(t)] - c(d + d_{eff}). \quad (17)$$

This can be solved numerically once we get $F_b(t)$.

## IV. ASYMPTOTIC BEHAVIOR AND SCALING LAW

Let us start from the following expression of $F_b(t)$,

$$F_b(t) - \bar{F}_b = \left(\frac{A}{B} - \bar{F}_b\right)(1 - \exp(-Bt)),$$

$$A \equiv c(d + d_{eff}), B \equiv q_b'(d + d_{eff}) + \mu, \mu \equiv r + a - g_b,$$

$$\text{with } \bar{F}_b = \frac{cd_{eff}}{q_b'd_{eff} + \mu} \text{ or } d_{eff} = \frac{\mu\bar{F}_b}{c - \bar{F}_bq_b'}. \quad (18)$$

Here we have taken $d_{eff}^{q_n} = d_{eff}^c \equiv d_{eff}$ for simplicity.

The function $F_b(t)$ takes the following form in the limit $Bt \gg 1$,

$$\lim_{Bt \to \infty}(F_b(t) - \bar{F}_b) = \hat{F}_b - \bar{F}_b. \quad (19)$$

Namely it tends to constant value $\hat{F}_b - \bar{F}_b$;

$$\hat{F}_b \equiv \frac{A}{B} = \frac{c(d + d_{eff})}{q_b'(d + d_{eff}) + \mu}. \quad (20)$$

On the other hand, for $Bt \ll 1$,

$$\lim_{Bt \to 0}(F_b(t) - \bar{F}_b) = (\hat{F}_b - \bar{F}_b)Bt. \quad (21)$$

This shows linear dependence on time t with its slope $k$ with the dose rate $d$ multiplied by some factor,

$$F_b(t) - \bar{F}_b = kt,$$
$$k \equiv A - B\bar{F}_b = c(d + d_{eff}) - [q'_b(d + d_{eff}) + \mu]\bar{F}_b$$
$$= [c - q'_b\bar{F}_b](d + d_{eff}) - \mu\bar{F}_b = [c - q'_b\bar{F}_b]d. \quad (22)$$

Thus the effective receptivity to radiation stimulus is a little bit reduced because of broken cell death effect.

This indicates that in the limit of starting period, i.e., for very short time after irradiation starts, the excess of broken cell number function behaves almost proportional to the time development.

Finally it turns out that $F_b(t)$ is found to be reduced to a kind of scaling behavior as follows, showing a simple scaling law,

$$\Phi(\tau) \equiv \frac{F_b(t) - \bar{F}_b}{\left\{\lim_{t \to \infty}(F_b(t) - \bar{F}_b)\right\}} = \frac{F_b(t) - \bar{F}_b}{\hat{F}_b - \bar{F}_b}$$
$$= 1 - \exp(-\tau), \tau \equiv Bt. \quad (23)$$

This indicates that if the constant dose rate is fixed, all the experimental data can be reduced into single curve of the above (Fig. 4) if we scale each data into the above scaling form.

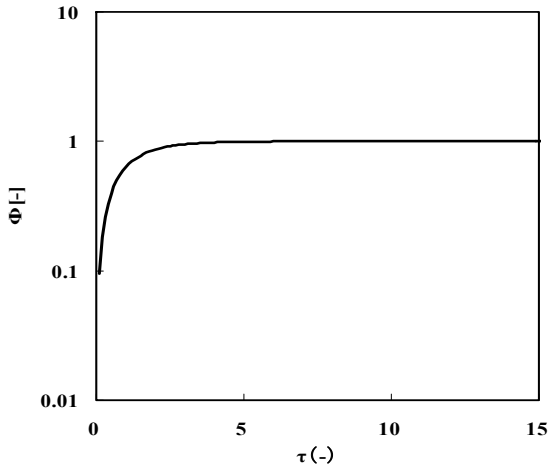

FIG. 4. Effective time dependence of $\Phi(\tau) = 1 - \exp(-\tau), \tau \equiv Bt$ of Eq. (23) indicates scaling law.

This can be checked from the experimental data performed by Russell groups as followings.

Let us take the famous accumulated data of mutation frequencies caused by irradiation in male mice. Such data is summarized by Russell and Kelly, where the data reported there were obtained by the so-called "mouse specific-locus test (SLT)" [6]. This uses seven visible markers and permits the detection of mutations involving any of the seven gene loci in the first-generation offspring of the irradiated parent. The results from treated and control adult males are to be compared in order to see the pure effects coming from artificial irradiation. The results are combined into a Table 1. The order of mutation frequency is of the order $10^{-5}$, so we can identify the experimental values of mutation frequency $f(t)$ to our number function $F_b(t)$, the solution of which is given,

$$f(t) = \left(\frac{c(d + d_{eff})}{\mu + q'_b(d + d_{eff})} - \bar{f}\right)\left(1 - \exp\left(-\left(\mu + q'_b(d + d_{eff})\right)t\right)\right) + \bar{f},$$
$$D(t) \equiv dt, \quad (24)$$

with $\bar{f} \equiv \frac{cd_{eff}}{\mu + q'_b d_{eff}}. \quad (25)$

TABLE I. Specific locus mutation rate data obtained in As spermatogonia by use of the seven-locus test stock provided by ref. 4

| Label | Category | Source of radiation | Total dose (Gy) | dose-rate (Gy/hour) | Mutations, no. | Offspring, no. | Mutation Frequency ×10⁵ per locus | Ref. |
|---|---|---|---|---|---|---|---|---|
| | Control | | | | 28 | 531,500 | 0.75 | 7 |
| | | | | | 11 | 157,421 | 1.00 | 8 |
| | | | | | 0 | 38,448 | 0.00 | 9 |
| A | Chronic | $^{137}$Cs | 3.00 | 0.00042 | 11 | 48,358 | 3.25 | 1 |
| B | | $^{60}$Co | 0.38 | 0.00060 | 7 | 79,364 | 1.26 | 8 |
| C | | $^{137}$Cs | 0.86 | 0.00060 | 6 | 59,810 | 1.43 | 7 |
| D | | $^{137}$Cs | 3.00 | 0.00060 | 15 | 49,569 | 4.32 | 7 |
| E | | $^{137}$Cs | 6.00 | 0.00060 | 22 | 53,380 | 5.89 | 7 |
| F | | $^{137}$Cs | 3.00 | 0.0030 | 24 | 84,831 | 4.04 | 1 |
| G | | $^{60}$Co | 6.71 | 0.0030 | 20 | 58,795 | 4.86 | 8 |
| H | | $^{60}$Co | 6.18 | 0.0048 | 5 | 22,682 | 3.15 | 8 |
| I | | $^{137}$Cs | 3.00 | 0.0054 | 10 | 58,457 | 2.44 | 7 |
| J | | $^{137}$Cs | 5.16 | 0.0054 | 5 | 26,325 | 2.71 | 7 |
| K | | $^{137}$Cs | 8.61 | 0.0054 | 12 | 24,281 | 7.06 | 7 |
| L | | $^{137}$Cs | 6.00 | 0.480 | 10 | 28,059 | 5.09 | 10 |
| M | Acute | X-ray | 3.00 | 54.000 | 40 | 65,548 | 8.72 | 7 |
| N | | X-ray | 6.00 | 54.000 | 111 | 119,326 | 13.29 | 7 |
| O | | X-ray | 6.70 | 43.200 | 12 | 11,138 | 15.39 | 8 |

The above function Eq. (24) includes only three parameters, $c$, $\mu$, $q'$. In a separate paper, we determine

RADIATION EXPOSURE THEORY COMPARISON OF DATA ON MUTATION FREQUENCIES OF MICE

them by using the chi-squared fit procedure so as to match the estimation for the observed data [13],

$$\mu = 3.13 \times 10^{-3} \text{ [1/hr]},$$
$$q'_b = 1.00 \times 10^{-1} \text{ [1/Gy]},$$
$$c = 2.91 \times 10^{-5} \text{ [1/Gy]}. \quad (26)$$

Together with the value $d_{eff}$ which is obtained the control data $\bar{f}$,

$$d_{eff} = \frac{\bar{f}\mu}{c - \bar{f}q'} = 1.11 \times 10^{-3} \text{ [Gy/hr]}, \quad (27)$$

we have quantities of all the necessary parameters.

Now that we have all the parameters of the above, we can straightforwardly calculate the numerical values of $\Phi(\tau)$ of Eq. (23). The plotted results are shown in Fig.5.

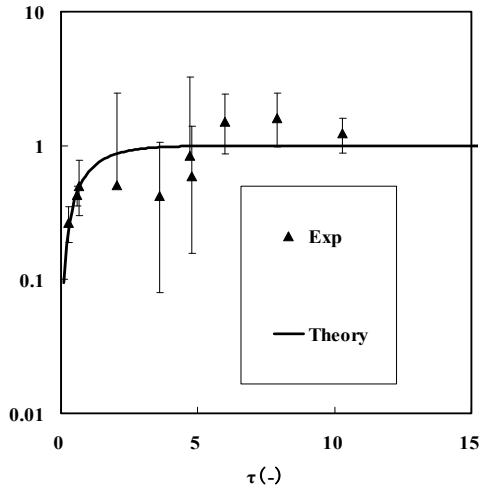

FIG. 5. Experimental data on the Renormalized number function $\Phi(\tau)$ in $(\tau, \Phi)$ plane.

It turns out that the scaling law predicted by RET works quite well to reproduce the realistic experimental data.

Fig.5 RET predicts the scaling law in which all the experimental data points lie on the unique line equation (see Eq. (23)).

$$\Phi(\text{mutation frequency} : \tau) = \frac{f(t) - \bar{f}}{\frac{A}{B} - \bar{f}}$$
$$= 1 - \exp(-\tau), \tau \equiv Bt. \quad (28)$$

## V. CONCLUSION AND DISCUSSION

We have shown that RET is a powerful tool to estimate biological damage caused by radiation. It reproduces reasonably well the famous experimental data on mutation induction in mice which were extensively carried out after the World War II. Moreover it predicts unique universal scaling function to which all the data points of mutation frequency reduce. Such scaling law exists for the data with various values of constant dose rate. The data summarized by Russell and Kelly are the appropriate set of data with a variety of dose rate and total dose [4].

Although the data of mutation frequency on the responses induced by the exposure to low levels of ionizing radiation is the most famous data obtained by extensive experiments, the remarkable coincidence of the prediction of RET encourages us to extend our analysis to various kinds of experiments performed today. RET may be a first good example to be applicable for the estimation of radiation risks.

We would like to stress that RET can apply to every kind of experiment with varying the time dependent dose rate experiments. For example, fractionation irradiation experiment provides us with further information to check RET validity.

We hope that RET may open the window to estimate the radiation risk in a quantitative way.

## ACKNOWLEDGEMENTS

The authors would like to thank to K. Uno, K. Fujita, M. Abe, H. Utsumi and O. Niwa for encouragement and informing us many experimental data of frequency of mutation for various irradiation exposure cases. Also thanks are due to Ichikawa for his collaboration in the analysis in early stage and the members of LDM, for their various comments and discussions.

RADIATION EXPOSURE THEORY COMPARISON OF DATA ON MUTATION FREQUENCIES OF MICE